\documentclass{IET-Conf-Paper}

\usepackage{graphicx}
\usepackage{physics}
\usepackage{algorithm}
\usepackage{algpseudocode}
\usepackage{caption}
\usepackage{subcaption}

\def\BibTeX{{\rm B\kern-.05em{\sc i\kern-.025em b}\kern-.08em
    T\kern-.1667em\lower.7ex\hbox{E}\kern-.125emX}}
\begin{document}

\title{ON QUANTUM ANNEALING WITHOUT A PHYSICAL QUANTUM ANNEALER}

\author{Ameya Bhave\ad{1}\corr, Ajinkya Borle\ad{1}}

\address{\add{1}{CSEE Department, University of Maryland Baltimore County, Baltimore, USA}
\email{ameyab1@umbc.edu}}


\begin{abstract}
Quantum annealing is an emerging metaheuristic used for solving combinatorial optimisation problems. However, hardware based physical quantum annealers are primarily limited to a single vendor. As an alternative, we can discretise the quantum annealing process (discretised quantum annealing or DiQA) and use it on gate-model quantum computers. In this work, we first benchmark DiQA against simulated annealing for a similar number of steps. We then propose and evaluate a hybrid quantum-classical heuristic: Quantum Accelerated Simulated Annealing (QASA), where the traditional classical annealing procedure can be sped up with the use of (relatively) low depth DiQA circuits. This is done by (i) running a partial annealing scheme with a fraction of the depth of the complete circuit (ii) sampling the results from the circuit and fitting a Gibbs distribution on it and (iii) Using the inverse temperature of the Gibbs distribution and the best sample to initialise simulated annealing (SA). Our simulation results show QASA performing comparably to SA but for a reduced amount of steps. With the promising results of our work, we hope to generate interest for potential future work in the area of fixed-parameter quantum optimisation.
\end{abstract}

\maketitle

\section{Introduction}
Combinatorial optimisation is one of the most promising use-cases for quantum computing and there have been many heuristics and approximate algorithms proposed for both near and far term quantum hardware. Among them, quantum annealing (QA) \cite{kadowaki1998quantum} is one of the most popular and mature metaheuristics for solving combinatorial optimisation problems with proposed applications in many domains from life-sciences, logistics, telecommunications and many more \cite{dwave_apps}. 

However, while classical Monte-Carlo simulations of quantum annealing is a popular and well-studied topic, the only vendor that successfully implemented this metaheuristic at the hardware level is D-wave\textsuperscript{TM} whose physical quantum annealers are analogue computers that implement QA as a physical process. In 2023, their quantum annealer was shown to be faster than simulated quantum annealing (SQA) for crystal dynamics \cite{king2023quantum}.

Since the gate-model of quantum computing is universal in nature \cite{deutsch1989quantum}, an alternative to hardware-based quantum annealers could be to implement quantum annealing on gate-model quantum computers \cite{willsch2020benchmarking}\cite{sack2021quantum}\cite{willsch2022gpu}. Broadly, we can call them discretised quantum annealing (DiQA) schemes since the model of computation considered here is not analogue. It should be noted that while different schemes to do the above have been proposed, the primary method for optimisation on gate-model quantum computers (QCs) have been variational quantum algorithms, such as the quantum approximate optimisation algorithm \cite{farhi2014quantum}.

While we acknowledge that a physical quantum annealer would probably outperform a DiQA scheme on gate-model QCs for most of the scenarios (due to the maturity, longer coherence times and application-specificity of the former's hardware); it is important to study DiQA as an alternative especially since the former is exclusive to only one commercial vendor. In this work, our aim is to do an exploration of DiQA to get insights on how it may be leveraged in the future to solve non-trivial problems. This means one of the focus areas is on using relatively low depth circuits\footnote{With respect to the depth of circuit calculated if the annealing schedule of a standard run of D-wave was to be properly converted. } as a subroutine in a larger optimisation heuristic. Our results can be divided into two parts.

First, we benchmark DiQA and simulated annealing for randomly generated Ising problems for variable sizes ranging from $N=10$ to $N=22$. The results show DiQA outperforming simulated annealing for a discretised annealing schedule of $200$ steps each. These would be the baseline results against which we compare other results in our work. 

We then propose and evaluate a hybrid quantum-classical heuristic for optimisation. It involves running a fraction of a DiQA circuit and sampling from it in the computational basis. Then we fit a Gibbs distribution on the observed bitstrings to obtain the inverse temperature $\beta$. Finally, the temperature $\beta$ and the best solution obtained from the distribution are supplied as initial parameters for simulated annealing. Initial evaluation against the results from the runs of non-hybrid SA suggests how such heuristics may be able to speed up classical techniques (with the help of a low depth quantum subroutine). 

 The rest of this paper is structured as follows: section \ref{sec:bnrw} covers the necessary background and related work, section \ref{sec:prelim_exp} covers the results of the set of initial experiments on DiQA and simulated annealing, section \ref{sec:QASA} is on our hybrid quantum-classical heuristic. We briefly discuss different possible directions that can be taken in the future in section \ref{sec:future_work} and we finish the paper with section \ref{sec:conclusion}: the conclusion.

\section{Background and Related Work}\label{sec:bnrw}
\subsection{Background}\label{sec:background}
Here, we describe the  necessary background and notations for this work.
\subsubsection{Quantum Computing} Quantum Computing can be defined as leveraging quantum mechanical phenomena like superposition, entanglement, and quantum measurement for computational purposes \cite{nielsen2002quantum}.  The smallest unit of information is a two level quantum state known as a qubit, which can be represented as
\begin{align}
    \ket{\psi} &= \mu\ket{0} + \nu\ket{1}\\
    \text{where }\ket{0} &= \begin{pmatrix}
    1 \\
    0 
    \end{pmatrix} \text{ and }\ket{1} = \begin{pmatrix}
    0 \\
    1 
    \end{pmatrix}
\end{align}
Where $\mu$ and $\nu$ are the amplitudes associated with $\ket{0}$ and $\ket{1}$ respectively, with $\abs{\mu}^2 + \abs{\nu}^2 = 1$. One of the most popular models of quantum computing is gate-model quantum computing, where quantum computing is represented in terms of discrete steps that manipulate a set of qubits (leveraging the above mentioned quantum phenomenon). We recommend the text by Nielsen and Chuang for a more detailed introduction of the subject \cite{nielsen2002quantum}.

\subsubsection{The (classical) Ising model}
This is a model from statistical mechanics, primarily used for modelling ferromagnetic materials, such as iron \cite{gallavotti1999statistical}. Computationally, it is an NP-Hard problem \cite{barahona1982computational} and in the recent years, the objective function associated with it has been used to represent hard problems from other domains \footnote{Not all NP-hard problems can be reduced to the Ising model. But all problems in NP can be reduced to the Ising model (some of them which are also NP-Hard), that too in polynomial time \cite{cormen2022introduction}.}  \cite{mcgeoch2013experimental,lucas2014ising}.

In the two-dimensional case, the energy or objective function of the classical Ising model can be described as follows
\begin{align}
    E(\sigma) &= \sum_{i}h_i\sigma_i + \sum_{i<j}J_{ij}\sigma_{i}\sigma_{j}\label{eq:ising}
\end{align}
Where $\sigma$ is a set of $N$ bipolar variables and $\sigma_i \in \{-1,+1\}, i = 1,2,\ldots N$. The primary goal here is to encode a combinatorial optimisation problem of interest as a minimisation or maximisation problem of the Ising objective function (mainly the former), send it to a quantum optimiser like the D-wave\textsuperscript{TM}'s physical quantum annealer, and then decode the results back to the original problem.

An equivalent objective function to Eqn(\ref{eq:ising}) is the quadratic unconstrained binary optimisation (QUBO) objective function, where the variables have binary values. Since our work deals only with Ising problems, we mention the above for the sake of completeness. But readers interested in this topic are recommended to read the book by Crama and Hammer \cite{crama2011boolean}.

\subsubsection{Quantum annealing (QA)}
Quantum annealing is a metaheuristic that attempts to find the groundstate (or the global minima) of a problem Hamiltonian\footnote{The matrix that describes the total energy of a physcial system}. The most well known problem Hamiltonian is the one for two-dimensional classical Ising Hamiltonian $\hat{H}_P$.
\begin{align}
    \hat{H}_{p} &= \sum_{i}h_i\hat{\sigma}_i^{(z)} + \sum_{i<j}J_{ij}\hat{\sigma}_{i}^{(z)} \hat{\sigma}_{j}^{(z)}\label{eq:hamp}\\
    \text{where } \hat{\sigma}_{i}^{(z)} &= (\otimes_{1}^{i-1}\hat{I}) \otimes (\hat{\sigma}^{(z)}) \otimes (\otimes_{i+1}^{N}\hat{I})\label{eq:sigmaz}
    \\ \text{ and } \hat{\sigma}^{(z)} &= \begin{pmatrix}
    1 & 0\\
    0 & -1
    \end{pmatrix}\label{eq:pauliz}
\end{align}
Eqn(\ref{eq:hamp}) is the Hamiltonian for the Ising energy function described in Eqn(\ref{eq:ising}). The process of quantum annealing at any given time can best be described by the following Hamiltonian
\begin{align}
    \hat{H}(s) &= -\frac{A(s)}{2}\hat{H}_{M} + \frac{B(s)}{2}\hat{H}_{P}\label{eq:anneal_hamil}\\
    \text{where }s &\in [0,1]    
\end{align}

Here, $s$ is the normalised time parameter. $A(s)$ and $B(s)$ are functions known together as the quantum annealing schedule. $\hat{H}_{M}$ is the starting Hamiltonian for which we can easily prepare the ground state. The most popular choice for this Hamiltonian is
\begin{align}
    \hat{H}_{M} &= \sum_{i}\hat{\sigma}^{(x)}_{i}\label{eq:hamm}\\
    \text{where } \sigma^{(x)} &= \begin{pmatrix}
    0 & 1\\
    1 & 0
    \end{pmatrix}\label{eq:paulix}
\end{align}

At $s=0$, all $N$ variables involved have an equal probability  of being either +1 or -1, $A(s)$ is very high and $B(s)\sim 0$. As $s$ goes from 0 to 1, $A(s)$ decreases and $B(s)$ increases, allowing the quantum state to go from the ground state of $\hat{H}_{M}$ to the ground state of $\hat{H}_{P}$ as long as the process is done slowly\footnote{Governed by the principle of adiabatic evolution \cite{farhi2000quantum}.}. Interested readers are recommended the paper by Kadowaki and Nishimori to understand this topic in depth \cite{kadowaki1998quantum}. This process is implemented either as a physical quantum annealer \cite{bunyk2014architectural} or as classical simulations \cite{martovnak2002quantum} with the former recently showing a speedup against the latter for crystal dynamics \cite{king2023quantum}.

\subsubsection{Quantum approximate optimisation algorithm (QAOA)}\label{sec:qaoa}
The most popular algorithm for combinatorial optimisation on the gate-model computer is the quantum approximate optimisation algorithm (QAOA) \cite{farhi2014quantum}. It involves the Hamiltonian simulations of $\hat{H}_{M}$ and $\hat{H}_{P}$ as
\begin{align}
    U(\hat{H}_M,\delta) &= e^{-i\delta\hat{H}_{M}} = \prod_{i}e^{-i\delta\sigma^{(x)}_{i}}\label{eq:unihm}\\
    U(\hat{H}_P,\gamma) &= e^{-i\gamma\hat{H}_{P}} = \prod_{i}e^{-i\gamma h_{i}\sigma^{(z)}_{i}} \prod_{i<j}e^{-i\gamma J_{ij}\sigma^{(z)}_{i}\sigma^{(z)}_{j}}\label{eq:unihp}
\end{align}
where $\delta \in [0,\pi]$ and $\gamma \in [0,2\pi]$ are angles\footnote{Typically represented as $\beta$ and $\gamma$, our change in notation is to avoid ambiguity with another value also represented by $\beta$.} applied in simulating the Hamiltonian. These Hamiltonian simulations can be performed by a combination of rotation and CNOT gates described below briefly
\begin{align}
         R_x(\omega) &= e^{-i\frac{\omega}{2}\hat{\sigma}^{(x)}} = 
    \begin{pmatrix}
    \cos{\omega/2} & -i \sin{\omega/2}\\
    -i \sin{\omega/2} & \cos{\omega/2}
    \end{pmatrix}\label{eq:rot_x}\\
    R_z(\omega) &= e^{-i\frac{\omega}{2}\hat{\sigma}^{(z)}} = 
    \begin{pmatrix}
    e^{-i \omega/2} & 0\\
    0 & e^{i \omega/2}
    \end{pmatrix}\label{eq:rot_z}\\
    \text{CNOT} &= 
    \begin{pmatrix}
    1 & 0 & 0 & 0\\
    0 & 1 & 0 & 0\\
    0 & 0 & 0 & 1\\
    0 & 0 & 1 & 0\label{eq:CNOT}\\
    \end{pmatrix}
\end{align}
For this algorithm, the system also begins in an equal superposition of all possible states 
\begin{align}
    \ket{\psi_{0}} = \frac{1}{\sqrt{2^N}}\bigl( \ket{0} + \ket{1} \bigr)^{\otimes N}\label{eq:qaoa_init}
\end{align}
A set of $p$ pairs of alternating Unitary operators are applied (based on equations \ref{eq:unihm} and \ref{eq:unihp}) for angles $\Delta = (\delta_{1},\delta_{2},...,\delta_{p})$ and $\Gamma= (\gamma_{1},\gamma_{2},...,\gamma_{p})$
\begin{align}
    \ket{\psi_{p}} &= U(\hat{H}_M,\delta_{p})U(\hat{H}_P,\gamma_{p})...U(\hat{H}_M,\delta_{1})U(\hat{H}_P,\gamma_{1})\ket{\psi_{0}}\\
    \text{or } \ket{\psi_{p}} &= U_{p}U_{p-1}...U_{2}U_{1}\ket{\psi_{0}}\\
    \text{where }
    U_{l} & = U(\hat{H}_M,\delta_{l})U(\hat{H}_P,\gamma_{l})\text{,  }
\end{align}
Here $l= 1,2,\ldots,p$. This parameterised circuit is known as the QAOA-ansatz and $p$ is known as QAOA-depth\footnote{Not to be confused with the related concept of gate depth. Each Unitary matrix from $U_{1}$ up to $U_{p}$ would have a fixed number of gates in it, depending on the problem and the QPU architecture.}. The task is to figure out the optimal values of these p angles that would maximise the probability of the ground state(s) to be measured\footnote{For converting the measured qubit into bipolar values, $\ket{0}\rightarrow+1$ and $\ket{1}\rightarrow-1$ (essentially the eigenstates and eigenvalues of a $\sigma^{(z)}$ matrix). }. This is typically done in a variational manner where the expectation value of the quantum state is calculated and the are parameters adjusted accordingly \cite{harrigan2021quantum}. In our work, we do not variationally update the parameters to find the best values. Instead, we use the procedure described next.

\subsubsection{Discretised Quantum Annealing (DiQA)} Like mentioned before, this is a general approach to express an analogue computational process like QA to convert it into discrete unitary operations. Our version of DiQA makes use of the second-order Suzuki-Trotter decomposition mentioned in the work by Willsch et al. \cite{willsch2022gpu}. Essentially, given a quantum schedule defined by the tuple of functions $\bigl(A(s),B(s)\bigr)$, we can discretise the QA process into $p$ steps of the QAOA-ansatz by the following equations:
\begin{align}
    \delta_{l} &= \frac{-\tau\bigl( A(s_{l+1}) + A(s_{l}) \bigr)}{2}, \hspace{1em} && l = 1,2,\ldots,p-1 \label{eq:del_l}\\
    \delta_{p} &= \frac{-\tau A(s_{p})}{2} && \label{del_p}\\
    \gamma_{l} &= B(s_{l}), \hspace{1em} && l = 1,2,\ldots,p \label{eq:gamma_l}
\end{align}

Here, the normalised time parameter is also discretized into $p$ steps. $\tau$ represents the amount of time taken per step. We use the annealing schedule for {\tt dwave\_2000Q\_6} machine \cite{dwave_schedule} in our experiments. These schedules are measured in Ghz\footnote{the original energies were divided by the Plancks constant $h$}
and so, for correctly converting these frequencies into angles, we need to multiply $A(s)$ and $B(s)$ with a factor of $1/4\pi$ each\footnote{ $1/2\pi$ for converting frequency into radians/sec and an additional $1/2$ because of Eqn(\ref{eq:anneal_hamil}).}. In the context of the D-wave\textsuperscript{TM}'s annealing schedule, $\tau$ is measured in nanoseconds. The value for $\tau$ we use for all our experiments down below is $0.8$ns.

\subsubsection{Simulated Annealing (SA)}
Simulated Annealing is a family of heuristics that aim to optimise a cost function by randomly sampling from the solution space of an objective function and then accepting or rejecting a new sample based on the current sample and a temperature parameter \cite{bertsimas1993simulated}. In the context of our work, the objective is to minimise Eqn(\ref{eq:ising}). The process starts with a inverse temperature parameter $\beta = \beta \sim 0$ and ends with $\beta = \beta_{max} \gg 0$. Specifically, for all our experiments, we use $\beta_{0} = 0.01$ and $\beta_{max} = 100$ with an exponential schedule given by the set $T_{inv}$.
\begin{align}
    T_{inv} = \{\beta_{i} \vert \beta_{i} = \beta_{0}\big(\frac{\beta_{max}}{\beta_{0}}\big)^{i/b} \text{ and } 0 \leq i \leq b-1, i \in \mathbf{Z} \}
\end{align}
where $b$ is the number of steps for which the schedule is defined. At each step of the annealing process, a number of monte carlo steps of sampling can be taken, governed by the $sweeps$ parameter.  Further details are presented in Algorithm \ref{alg:SA}.

\begin{algorithm}
\caption{Basic Simulated Annealing(pseudocode)}\label{alg:SA}
\begin{algorithmic}[1]
\Procedure{MAIN}{$h,J,sweeps,T_{inv}$}
\State Randomly initialise all Ising spins $\sigma_{i} \in \{-1,+1\}$, $1 \leq i \leq N$
\Repeat
    \State Get next $\beta$ from $T_{inv}$ 
    \For{ k from 1 to $sweeps$}
        \State Select a variable $i$ randomly
        \State Update $\sigma \leftarrow$ MCSTEP($\sigma,i,h,J,\beta$)
    \EndFor
\Until{all $\beta$s in $T_{inv}$ are covered}
\State \textbf{return} $\sigma$
\EndProcedure
\Procedure{MCSTEP}
{$\sigma,i,h,J,\beta$}

\State Calculate current state energy $E_{curr}$ with Eqn(\ref{eq:ising})
\State Flip sign $\sigma_{i} \leftarrow -\sigma_{i} $ and calculate the proposed state energy $E_{prop}$
\State Calculate $E_{diff} = E_{prop} - E_{curr}$
\If{$E_{diff} < 0 $} 
    \State \textbf{return} $\sigma$ \Comment{Accept move}
\Else
    \State Randomly sample a value $r$ from $U(0,1)$
    \If{$r < e^{-\beta E_{diff}}$}
        \State \textbf{return} $\sigma$ \Comment{Accept this move as well}
    \Else
        \State Flip $\sigma_{i} \leftarrow -\sigma_{i}$ \Comment{Reject move, revert back}
        \State \textbf{return} $\sigma$ \Comment{Return old $\sigma$}
    \EndIf
\EndIf
\EndProcedure
\end{algorithmic}
\end{algorithm}

\subsubsection{Gibbs distribution}
A Gibbs or Boltzmann distribution is a probability distribution that is defined using an energy function and the inverse temperature parameter $\beta$. 
\begin{align}
    P(\sigma) &= \frac{e^{-\beta E(\sigma)}}{Z}\label{eq:gibbs_fit}\\
    \text{where } Z &= \sum_{\{\sigma\}} e^{-\beta E(\sigma)}\label{eq:Z}
\end{align}
Where Eqn(\ref{eq:Z}) describes the partition function, which is intractable to calculate in most cases. In our case, the variables are bipolar and the energy function is defined by Eqn(\ref{eq:ising}).
\subsection{Related work}\label{sec:related_work}
As mentioned above, there have been various works that have discretised quantum annealing for circuit based quantum computers, predominantly for the QAOA-ansatz \cite{willsch2020benchmarking}\cite{sack2021quantum}\cite{willsch2022gpu}. In our work, we use DiQA (i) to assess its performance against simulated annealing and (ii) as a subroutine to speed up simulated annealing. 

At a glance, the work on Quantum-enhanced Markov chain Monte Carlo (QE-MCMC) \cite{layden2023quantum} seems to be similar to our proposed QASA technique. There are however a few differences: (i) QE-MCMC is a generalised framework focused on sampling a distribution whereas with QASA our end goal is to get the best solution for a cost Hamiltonian, (ii) QE-MCMC is very broad in its scope of potential applications whereas  QASA is more focused on solving combinatorial optimisation problems (iii) QE-MCMC is an iterative algorithm that switches between the quantum and the classical step (with a feedback to the quantum step), whereas QASA does not loop between its quantum and classical steps. We recommend interested readers the work on QE-MCMC by Layden et al. \cite{layden2023quantum} as well. Another directly related work to QASA is the work that uses the D-wave\textsuperscript{TM}  physical quantum annealer and neural networks for accelerating MCMC on the Ising model \cite{scriva2022accelerating}.

Our QASA heuristic is inspired by the work of França and Patron \cite{stilck2021limitations}. In that work, the authors estimate the quality of the results of a quantum optimiser by fitting its outputs to a Gibbs distribution and obtaining the corresponding $\beta$ (larger the value, the better).  In this work, we use a similar idea, but use the $\beta$ and a bipolar bitstring $\sigma$ from DiQA with (one of) the lowest energy as the starting parameters for SA.

It is also important to mention the work done on neutral-atom quantum computers that can do analogue or dual analogue/discrete quantum computing. In brief, they have the ability to perform analogue Hamiltonian evolution for applications in combinatorial optimisation \cite{scholl2021quantum}\cite{nguyen2023quantum}. Though our demonstration of QASA is with DiQA, the basic procedure would also be able to incorporate analogue quantum computing as well.

\section{Preliminary Experiments}\label{sec:prelim_exp}
\subsection{Experiment setup}\label{sec:prelim_exp_setup}
We generated $100$ random Ising problems with a 6-regular graph interactivity\footnote{The justication being that the first quantum annealers were based on the chimera graph, a structure with an average degree of six \cite{mcgeoch2013experimental}.} between the variables for sizes $N = \{2i\vert 5 \leq i \leq 11, i\in\mathbf{Z}\}$. The coefficients are uniformly sampled in the range $h_{i}\in [-2,2]$ and $J_{ij}\in [-1,1]$ rounded to one decimal place. For the purposes of the experiments, we found the groundstate(s) for each problem by brute-forcing every possible combination of variable values. This is important since the probability of successfully measuring (or reaching) a ground state $\mathcal{P}$ is the key output metric for us.

We ran DiQA simulations with a QAOA-ansatz depth of $p=100$ and $p=200$ against SA with the same number of steps $b=p$ as above. For our SA runs, we also ran them for two settings (i) one sweep per $\beta$ and (ii) $N$ sweeps per $\beta$ for both $b=100$ and $b=200$ steps. All quantum simulations in this work were conducted using noiseless simulators with all-to-all qubit connectivity using the QISKIT package for the Python programming language  \cite{aleksandrowicz2019qiskit} and computers with standard {\tt x86-64} processors. The success probability $\mathcal{P}$ was calculated with the help of the resultant statevectors for DiQA. This is done to get an accurate estimate of what $\mathcal{P}$ would be in ideal conditions. For SA however, there is no direct equivalent of a statevector that we can keep track of, thus we need to approximate the success probability. We do this by running each problem 2000 times, counting the number of times we attain a ground state (at the end of a run) and dividing it by 2000.

\subsection{Results and discussion}\label{sec:prelim_exp_rnd}
\begin{figure}
    \centering
    \includegraphics[width=8.5cm, height=6cm]{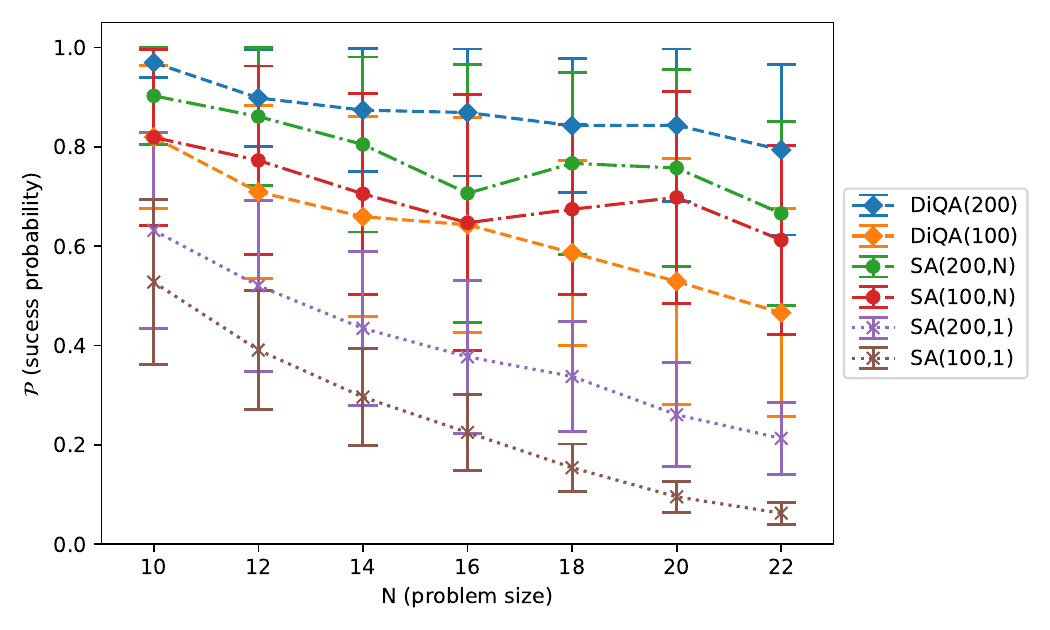}
    \caption{Success Probability $\mathcal{P}$ with respect to the problem size $N$ for different heuristics and parameters. The parameter denoted for DiQA is the circuit depth $p$ for which it was run. For SA, the first parameter indicates the number of annealing steps taken, and the second indicates the number of sweeps per anneal step. }
    \label{fig:figure_1}
\end{figure}
The results shown in figure \ref{fig:figure_1} show the success probabilities $\mathcal{P}$ as the number of variables $N$ increase. All points on the plot are median values of the $100$ problems and the error bars are  median absolute deviations (MAD). The general trend of $\mathcal{P}$ reducing as problem size $N$ increases is to be expected, since the heuristics were run with fixed (for DiQA and SA) or linearly increasing resources (for SA only).

The worst performing heuristic here is simulated annealing when its allowed to only take one sweep per $\beta$, even though the number of steps taken by it was similar to DiQA. When we increase the number of sweeps to $N$ per $\beta$ we can see SA with 100 steps and N sweeps perform better than DiQA with $p=100$. However, we see DiQA with $p=200$ marginally outperforming SA with $b=200$ steps and N sweeps. The improvement in performance becomes more significant when considering the run-time resources increase linearly for SA (for N sweeps) but are fixed for DiQA.

The results of these preliminary experiments would be useful for helping us with understanding the performance of our QASA hybrid quantum-classical heuristic. While previous comparisons between quantum annealing and SA have existed \cite{crosson2016simulated}\cite{djidjev2018efficient}\cite{streif2019comparison}, comparing a discretised version of quantum annealing (meant for gate-model quantum computers) with SA is still a useful result for the community. However, it is also important to keep in mind that these results shouldn't be taken as a proof of quantum advantage over classical annealing.
\section{Quantum Accelerated Simulated Annealing (QASA)}\label{sec:QASA}
\begin{figure*}[h]
    \centering
    \includegraphics[width=14cm, height=7cm]{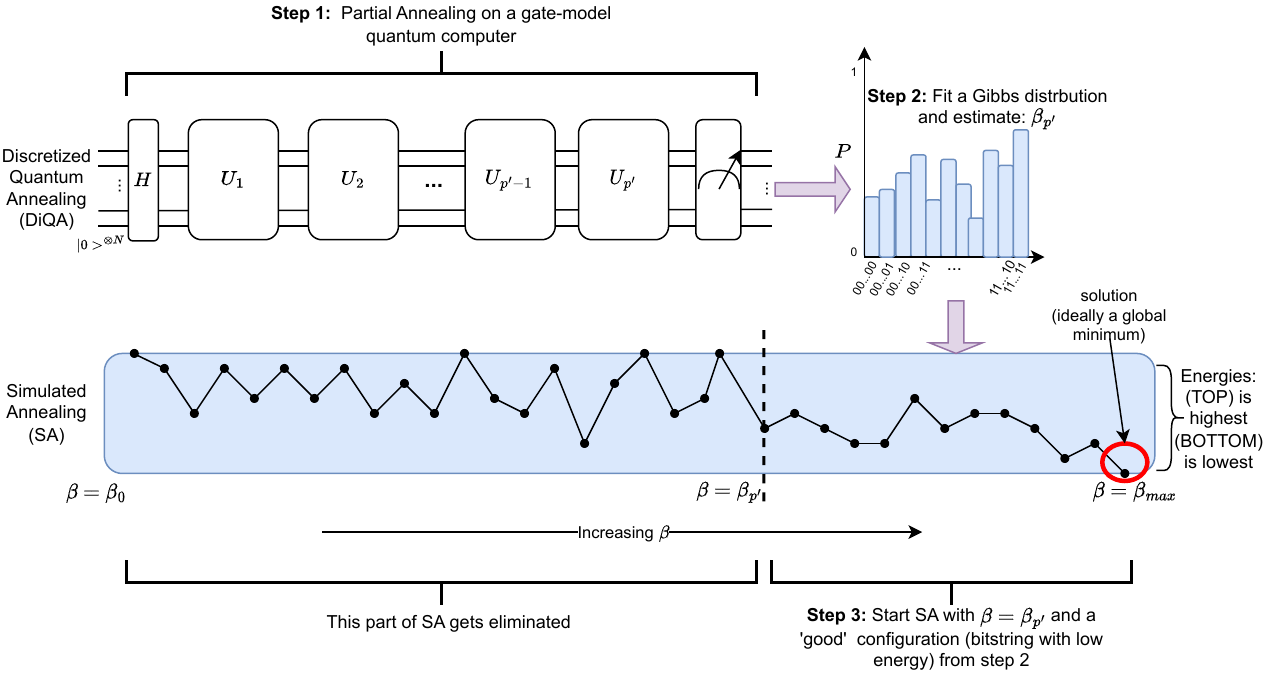}
    \caption{Block diagram showing the workflow of QASA. The quantum circuit in step 1 starts with applying hadamard gates on individual qubits (all initialised to $\ket{0}$), followed by Unitary operators up to $p'$ depth. The Y axis of the plot (not to scale) for step 2 shows the probability of each measured bit string. The illustration of the SA block is aimed at showing a basic classical annealing procedure that starts at a small inverse temperature $\beta_{0}$ and ends at $\beta_{max}$, where $\beta_{p'}$ lies somewhere in between the process. The line inside the SA block is a basic illustration of how the energy values of the system may vary as the heuristic progresses.}
    \label{fig:figure_2}
\end{figure*}

Our hybrid classical-quantum heuristic is aimed at leveraging a lower depth quantum circuit (relative to the depth of a standard DiQA run) to help with combinatorial optimisation tasks. Figure \ref{fig:figure_2} shows this process pictorially. But a detailed description of the process is given next.

\subsection{Description}\label{sec:QASA_desc}

\subsubsection{Partial annealing with DiQA}
In the very first step, a QA schedule is discretised into $p$ steps. The first $p'$ steps are chosen ($p' \leq p$) and a quantum circuit is made according to the rules of DiQA. At the end of the circuit we measure the outputs and get the probability distribution of measuring a particular bitstring back.
\subsubsection{Calculate the inverse temperature $\beta_{p'}$}
In this step, the $\beta_{p'}$ value will be calculated by fitting the probabilities from the step before as a Gibbs distribution as described in Eqn(\ref{eq:gibbs_fit}). However, since the partition function $Z$ is intractable to calculate in most cases, we can eliminate it from the equation by dividing Eqn(\ref{eq:gibbs_fit}) for two Ising configurations\footnote{A trick borrowed from SA itself \cite{bertsimas1993simulated}.}: $\sigma^{(1)}$ and $\sigma^{(2)}$\footnote{Here $\sigma_{(1)}$ is a set of $N$ bipolar values, essentially (one of) the Ising configuration for all $N$ variables. Not to be confused with $\sigma_{1}$ which holds the bipolar value (scalar) of the first variable.}
\begin{align}
\frac{P(\sigma^{(1)})}{P(\sigma^{(2)})} &= \frac{e^{-\beta_{p'}E(\sigma^{(1)})}}{e^{-\beta_{p'}E(\sigma^{(2)})}}
\end{align}
Taking the natural logarithm on both sides, we get
\begin{align}
    \ln{P(\sigma^{(1)})} - \ln{P(\sigma^{(2)})} &= \ln{e^{-\beta_{p'}E(\sigma^{(1)})}} - \ln{e^{-\beta_{p'}E(\sigma^{(2)})}}\\
    \text{or, }\beta_{p'} &= \frac{\ln{P(\sigma^{(1)})} - \ln{P(\sigma^{(2)})}}{E(\sigma^{(2)}) - E(\sigma^{(1)})}\label{eq:betap}
\end{align}
Thus, if we have two bitstrings  $\bigl( \sigma^{(1)},\sigma^{(2)} \bigr)$ with significant observable probabilities, we can use them to calculate $\beta_{p'}$. In the ideal scenario, the probability distribution is a native Gibbs distribution and so any two pairs of bitstring could be used to calculate the inverse temperature. However, when this is not the case, different bitstring pairs can produce different $\beta$ values. In this case, we recommend taking the average or median from all calculated $\beta$s. We shall see a demonstration of this in our experiments below.

\subsubsection{Run (shortened) simulated annealing}
With $\beta_{p'}$ and the bitstring for lowest energy found so far, we can now run SA for a shortened amount of steps up to $\beta_{max}$ (assuming $\beta_{p'} < \beta_{max}$). The DiQA procedure done prior to this is aimed at reducing the number of steps that SA needs to take (while not incurring a large overhead in itself).

\subsection{Experiment setup}\label{sec:qasa_exp_setup}
For experiments to evaluate QASA, the 700 Ising problems described in section \ref{sec:prelim_exp_setup} were used again. With this, we can  compare and contrast the results from the two different experiments. 

The annealing schedule of {\tt dwave\_2000Q\_6} was discretised for $p=200$ steps. From that, we ran DiQA circuits for the first $p'=50$ and $p'=100$ steps\footnote{We also ran it for $p'=150$ steps, but only the results from the prior two were used in the full QASA heuristic.}.

Since the probability distribution we receive from our partial runs of DiQA may not be native Gibbs distributions. In order to calculate $\beta_{p'}$ for each problem, we use an approximation method. This is done by calculating $m=50$ bitstrings with the largest probabilities. But since our experiments were done on a small $N$  size, it is generally likely that the largest probability configuration is the ground state. This would make rest of the computation trivial. Thus, for the purposes of this particular experiment, we purposely picked the second-largest probability\footnote{In order to show at least a modicum of robustness in QASA.}  bitstring as $\sigma^{1}$ and pair it with all the other $m-1$ bitstrings individually to make $m-1$ $\bigl( \sigma^{(1)},\sigma^{(2)} \bigr)$ tuples. Finally, we calculate the $\beta_{p}$ value for each of these tuples by Eqn(\ref{eq:betap}) and take the median value to represent the $\beta_{p'}$ for a problem.

The SA schedule $T_{inv}$ for $b=200$ steps is used here. Based on the closest value on the SA annealing schedule to $\beta_{p'}$, we calculate the starting step number $b'$ in the schedule. We use $N$ sweeps per step in this truncated SA procedure.

\subsection{Results and discussion}\label{sec:QASA_exp_rnd}

\begin{figure}
    \centering
    \includegraphics[width=8.5cm, height=6cm]{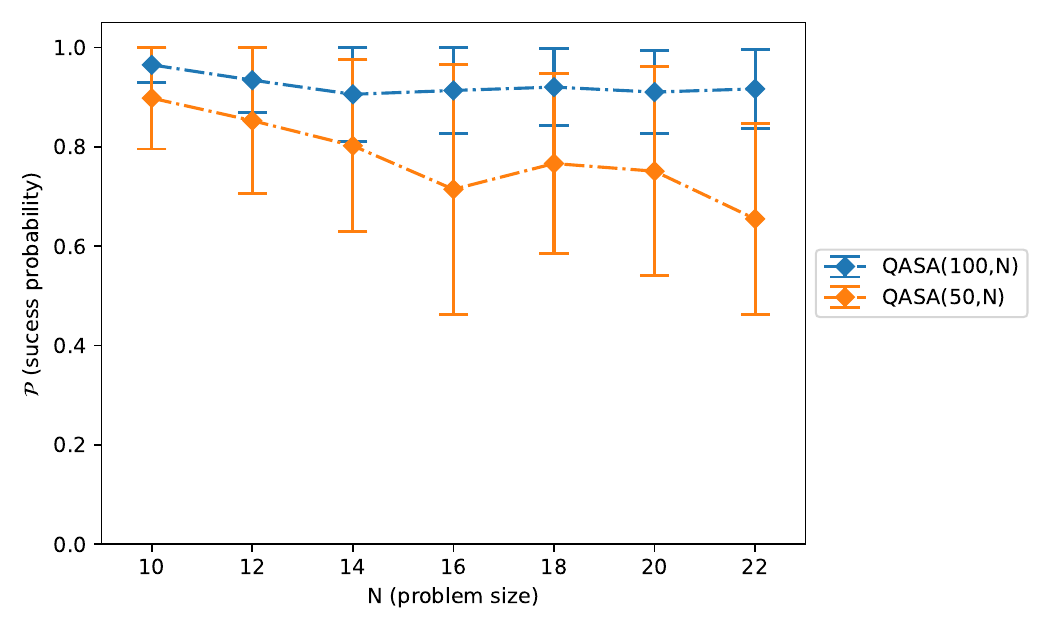}
    \caption{Success probability $\mathcal{P}$ for QASA when DiQA is run with $p'=100$ and $p'=50$, followed by SA for N sweeps per step (labelled as QASA(100,N) and QASA(50,N), respectively). Here the SA process is discretized into $b=200$ anneal steps but we skip the first $b'-1$ steps based on $\beta_{p'}$ which is calculated at the end of DiQA. }
    \label{fig:figure_3}
\end{figure}

The results of our QASA experiments in figure \ref{fig:figure_3} show QASA with $p'=50$ perform similarly to SA\footnote{SA is run for $N$ sweeps per step in both cases} with $b=200$ in figure \ref{fig:figure_1}. For $p'=100$, we see the results from QASA being the best across all of the tests for the set of general-case problems (created in section \ref{sec:prelim_exp}). While these results are not conclusive on their own, they do show that a technique like QASA can help in optimisation tasks in the near to medium term.

\begin{figure}
  \centering
  \begin{subfigure}{0.5\textwidth}
   \includegraphics[width=8.5cm, height=6cm]{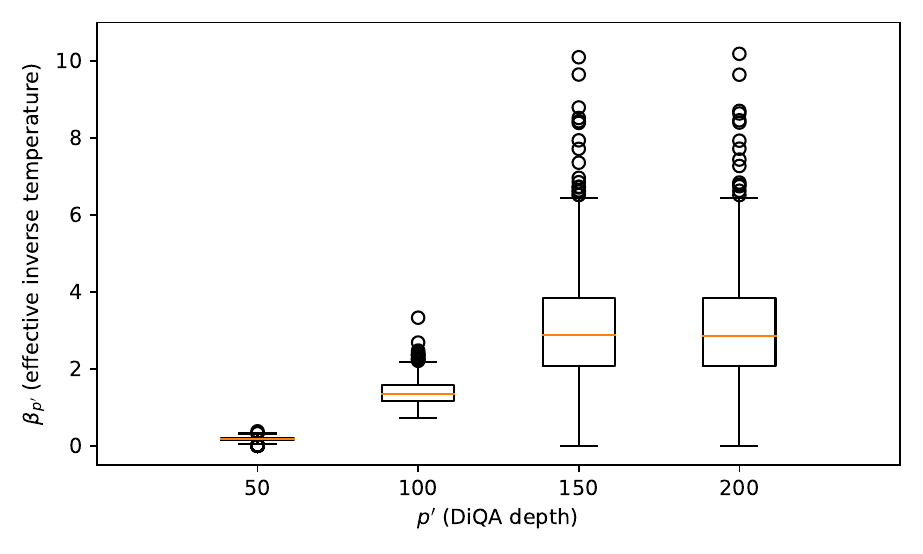}  
  \caption{ }
  \end{subfigure}
  \begin{subfigure}{0.5\textwidth}
  \includegraphics[width=8.5cm, height=6cm]{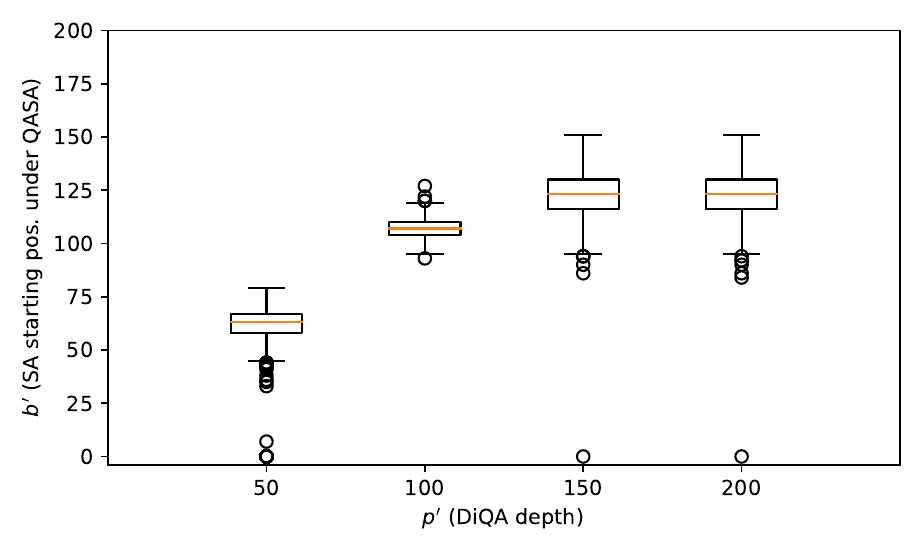}  
  \caption{ }
  \end{subfigure}
\caption{Boxplots depicting the spread of the $\beta_{p'}$ values (a) and the starting point $b'$ for SA under QASA (b) for $p'=\{50,100,150,200\}$. The values of $\beta_{p'}$ are approximated as mentioned in section \ref{sec:qasa_exp_setup} and the starting positions $b'$ are calculated with respect to the exponential SA schedule.}
\label{fig:figure_4}
\end{figure}

Like mentioned previously, although we ran the complete QASA heuristic for $p'=50$ and $p'=100$, we calculated $\beta_{p'}$ for $p'=\{50,100,150,200\}$ to get information about how different $p'$ values affect the inverse temperatures. Figure \ref{fig:figure_4} shows the boxplots of $\beta_{p'}$ and the starting position (or step) $b'$ for SA. We can see that although the median $\beta_{p'}$ for $p'=50$ and $p'=100$ is relatively close to zero ($\approx0.1815$ and $\approx1.34812$ respectively), the median $b'$ is 64 and 108 due to the exponential SA schedule. Another effect of the exponential schedule can be seen by the outliers for the boxplots of $p'=150$ and $p'=200$; for $\beta_{p'}$  they appear on top, but they appear on the bottom in the plot for $b'$.

It must be mentioned that since SA needs multiple Monte-carlo steps (or sweeps) per anneal step, the improvement for DiQA $p'=50$ and $100$ is much larger than 14 and 8 anneal steps respectively. Especially since  DiQA can employ parallelism of quantum gates for implementing much of its Unitary operator $U_{l}$ per anneal step. An in-depth analysis of the actual advantage in time saved would also depend on a quantum processing unit's (QPU) (i) qubit connectivity and its ability to do (ii) single control multiple target CNOT gates. Likewise, even though running the DiQA circuit for $p'=150$ only gives us a starting position of $b'=124$, it can still have a time advantage because of how differently an (efficient) anneal step needs to be implemented on a quantum circuit as opposed to a classical computer. It should be noted that at $p'=200$, figure \ref{fig:figure_4} stops showing any improvement at all (in $\beta_{p'}$ or $b'$). This can be due to several factors including (i) the highly variable results at higher depths as suggested by the error bars in figure \ref{fig:figure_1}, (ii) the limitations of our approximation technique to ascertain $\beta_{p'}$, (iii) the diminishing returns at higher $p$ of a QAOA-ansatz \cite{borle2021quantum}, among others. 

Now that preliminary testing for this technique has been done in this experiment, it is important to note that the approximation of $\beta_{p'}$ will be more noisy in practise due to many factors, including (i) shot-noise from sampling (ii) noise from the quantum computers. We shall discuss the latter in the next section on the possible future work.
\subsection{Feasibility for use on actual quantum devices: a brief comment}
One advantage of using the QAOA-ansatz is the prevalence  of work that has been done to evaluate its performance on real QPUs and noise-models. However, there still have not been many works that study the ansatz with a larger $p$.

The 2021 work by Harrigan et al. deals with variational QAOA on a superconducting-qubits based device and suggests a drop off in performance after $p=5$ (for their machine) \cite{harrigan2021quantum}. However, with the emergence of error-suppression and mitigation techniques, significant improvements in the results have been observed \cite{mundada2022experimental}. Another  major positive result is the work done by Kim et al. \cite{kim2023evidence}, which although is not about QAOA, does deal with implementing a larger number CNOT gates for a significant depth of a circuit successfully.

On trapped-ion QPUs, A simulation of QAOA for 40 qubits showed that that the `QAOA
performance does not degrade significantly as they scaled up the system size' \cite{pagano2020quantum}. One of the latest works\footnote{As of the time of writing.} has implemented advanced mixers for QAOA on a trapped-ion computer for $p>1$ successfully \cite{zhu2022multi}. This indicates a possibility of reducing DiQA depth by using better mixer Hamiltonians for the annealing evolution.

It appears that current QPUs may not be able to implement DiQA for a non-trivial depth size. But based on the latest work on error suppression and mitigation, DiQA may be feasible for implementation in the intermediate term (within 2 years \cite{castelvecchi2023ibm}, at the earliest), especially if advance mixers are used (for circuit depth reduction).

Like mentioned in section \ref{sec:related_work}, one of the most promising quantum platform for combinatorial optimisation is the neutral-atom quantum computer where the native problem Hamiltonian is slightly different to an Ising model \cite{nguyen2023quantum}. Nonetheless, adapting QASA for neutral-atom computers can be an interesting research direction.

\section{Future Work}\label{sec:future_work}
One of the most important work to be done in the future is to study DiQA and QASA with mixers more advanced than Eqn(\ref{eq:hamm}). This will give us an idea about the reduction in $p$. As mentioned above, A version of QASA for neutral-atom  quantum computers can be another interesting research direction.

Another interesting research direction to pursue would be to replace simulated annealing (SA) in QASA with simulated quantum annealing (SQA) \cite{martovnak2002quantum} that takes the same quantum annealing schedule as DiQA. In this approach, we would not need to calculate $\beta$ and can directly feed the quantum annealing parameters equivalent to the $p'+1$ step. Of course, this approach will only work if the mixer Hamiltonian is described by Eqn(\ref{eq:hamm})\footnote{Meaning if the annealing process is described by Stoquatic Hamiltonians.}.

And finally, for all future experiments, work should focus on better noise-resistant techniques to assess $\beta_{p'}$ in QASA. Performance metrics that allow for approximate answers, like approximation ratios \cite{farhi2014quantum} and time-to-solution \cite{king2015benchmarking} should also be considered. This is useful since a lot of applications of combinatorial optimisation can often suffice with a solution that may be a local minima with a sufficiently low energy rather than the global minima itself.

\section{Conclusion}\label{sec:conclusion}
In this work, we did a preliminary evaluation of the usefulness of discretised quantum annealing (DiQA) on quantum circuits, primarily as an alternative to a hardware based quantum annealer. Our  benchmark against simulated annealing showed better results for equal number of discretised anneal steps. Our main contribution was the hybrid quantum-classical heuristic that combined a (relatively) low depth DiQA circuit and simulated annealing with better informed parameters. Our experiments show QASA performing comparably to SA but for a reduced number of total steps (even with the overhead of DiQA).

Techniques like DiQA and its variants are not new, but they are comparatively less studied (as opposed to variational quantum algorithms). Our motivation with this work is to provide new perspectives to the research community, and to encourage more research on fixed-parameter quantum optimisation techniques, and hybrid quantum-classical heuristics that can incorporate them.
\section{Acknowledgements}
The authors would like to thank Dr Charles Nicholas and the DREAM laboratory in the University of Maryland Baltimore County (UMBC), for providing access to their high performance compute server.

\section{References}









\bibliographystyle{ieeetr}
\bibliography{references}
\end{document}